\documentclass[%
 reprint,
superscriptaddress,
 amsmath,amssymb,
 aps,
prc,
]{revtex4-2}
\usepackage{graphicx}
\usepackage{dcolumn}

\usepackage{bm}
\usepackage{color}

\usepackage{multirow}
\usepackage{mathtools}
\usepackage[utf8]{inputenc}

\begin{document}

 \preprint{APS/123-QED}

\title{A relativistic approach for determination of nuclear and neutron star properties in consideration of PREX-II results}
\author{Virender Thakur}
 \email{virenthakur2154@gmail.com}
 \affiliation{Department of Physics, Himachal Pradesh University, Shimla-171005, India}
 \author{Raj Kumar}
  \email{raj.phy@gmail.com}%
\affiliation{Department of Physics, Himachal Pradesh University, Shimla-171005, India}
\author{Pankaj Kumar}
 \affiliation{Department of Applied Sciences, CGC College of Engineering, Landran, Mohali 140307, India}
\author{Mukul Kumar}
\affiliation{Department of Physics, Himachal Pradesh University, Shimla-171005, India}
\author{C. Mondal}
\affiliation{Laboratoire de Physique Corpusculaire, CNRS, ENSICAEN, UMR6534, Université de Caen Normandie,
F-14000, Caen Cedex, France}
\author{Kaixuan Huang}
\affiliation{School of Physics, Nankai University, Tianjin 300071, China}
\author{Jinniu Hu}
\affiliation{School of Physics, Nankai University, Tianjin 300071, China}
\author{B.K. Agrawal}
\email{sinp.bijay@gmail.com}
\affiliation{Saha Institute of Nuclear Physics, 1/AF Bidhannagar, Kolkata 700064, India}
\author{Shashi K. Dhiman}
 \email{shashi.dhiman@gmail.com}
\affiliation{Department of Physics, Himachal Pradesh University, Shimla-171005, India}
\affiliation{School of Applied Sciences, Himachal Pradesh Technical University, Hamirpur-177001, India}

\begin{abstract}
 The bulk properties of   nuclear matter and neutron stars with the newly generated relativistic interaction DBHP are investigated which provides an opportunity to modify the coupling parameters keeping in view the finite nuclei,  nuclear matter, PREX-II data for neutron skin thickness in $^{208}$Pb and astrophysical constraints. The relativistic interaction has been generated by including all possible self and mixed interactions between $\sigma$, $\omega$, and $\rho$-meson up to the quartic order satisfying the naturalness behavior of parameters.  A covariance analysis is performed to assess the statistical uncertainties on the model parameters and observables of interest along with correlations amongst them.  We obtained a value of neutron skin thickness for $^{208}$Pb nucleus $\Delta r_{np}$ = 0.24 $\pm$ 0.02 fm.  The maximum gravitational mass of neutron star  and radius corresponding to the canonical mass  ($R_{1.4}$)  come out to be  2.03 $\pm$ 0.04 M$\odot$  and  13.39 $\pm$ 0.41  km respectively. The dimensionless tidal deformability,
${\Lambda}$ for a neutron star is also analyzed.
\end{abstract}
\maketitle


\section{INTRODUCTION}
 Neutron stars  (NSs) are  highly dense and asymmetric nuclear systems having a central density about 5-6 times the nuclear saturation density \cite{Lattimer2004}. The  studies  of the
NSs proclaim that their  internal structure are   quite  complex as new degrees of freedom like hyperons and quarks  may appear in the core. The NS properties   like  mass, radius, and tidal deformability can be  estimated using  equations of state  (EoSs) obtained within  various theoretical models \cite{Horowitz2001,Reed2021,Fattoyev2018}. One of such models is based on the relativistic interaction  which describes the interaction between nucleons through $\sigma$, $\omega$ and $\rho$  mesons. There are  several models of relativistic mean field (RMF) effective lagrangian density consisting of nonlinear $\sigma$, $\omega$, and $\rho$ terms and cross terms that have been analyzed for nucleonic  and  hyperonic matter and confronted  with the constraints of nuclear matter properties and astrophysical observations of NS masses \cite{Dhiman2007,Pradhan2022,Virender2022a,Virender2022b,Suman2022}.\\
 The nuclear theory studies \cite{Haensel2007,Lattimer2014,Baym2018} are mainly focusing on understanding the dense matter in NS. The constraints on EOS at high density are imposed with currently available lower bound  on  neutron star's maximum mass and radius \cite{Hebeler2010,Hebeler2013,Lattimer2012}. The precise measurement of masses of millisecond pulsars such as PSR J1614-2230 \cite{Demorest2010}, PSR J0348+0432 \cite{Antoniadis2013} show that the maximum mass of the NS  should be around 2 M$\odot$.
The recent observations with LIGO and Virgo of GW170817 event \cite{Abbott2018,Abbott2019} of Binary Neutron Stars merger
and the discovery of NS  with masses around 2$M_\odot$ \cite{Demorest2010,Antoniadis2013,Arzoumanian2018,Miller2019,Riley2019,Raaijmakers2019} have intensified the interest in these intriguing objects. The analysis of GW170817 has demonstrated the potential
of gravitational wave (GW) observations to yield new information relating to 
the limits on NS  tidal deformability.
The Lead Radius Experiment (PREX-II) has recently provided  a model-independent extraction of neutron skin thickness of $^{208}$Pb as $\Delta r_{np}$ = 0.283 $\pm$ 0.071 fm \cite{Adhikari2021}.  The $\Delta r_{np}$ has been identified as an ideal probe for the density dependence of symmetry energy - a key but poorly known quantity that describes the isospin dependence of the EOS of asymmetric nuclear matter and plays a critical role in various issues in nuclear physics and astrophysics. The neutron skin thickness of the Lead nucleus exhibit a strong positive linear correlation with the slope  of symmetry energy parameter (L) at saturation density.  The parameter $L$  that determines  the density  dependence of symmetry energy strongly affects the Mass-Radius relation and tidal deformability ($\Lambda$) of a neutron star and provides a unique bridge between atomic nuclei and neutron star \cite{Lattimer2001} . The large value of  $\Delta r_{np}$ = 0.283 $\pm$ 0.071 fm suggests a  large value of $L$   which yields a  very stiff EOS. This   generally gives rise to a large value of neutron star radius and the tidal deformability \cite{Reed2021}. The upper limit on $\Lambda_{1.4}$ $\leq$ 580 for GW170817 requires softer EOS and hence softer symmetry energy coefficient \cite{Abbott2018}. The heaviest observed  neutron star $M_{max}= 2.35\pm 0.17 M_\odot$ for the black-widow pulsar PSR J0952-0607 \cite{Romani2022} may place  stringent constraints on the symmetry energy at  high densities, since,   the EOS of symmetric nuclear matter (SNM) from heavy ion collisions flow data   \cite{Danielewicz2002} which is relatively soft and  limits the NS maximum mass.\\

The motivation of the present work is to generate a new parametrization of the RMF model which can accommodate  the properties of NSs within the astrophysical observations without compromising the finite nuclei  properties.
The RMF model used in the present work,  includes all possible self and mixed-coupling terms for the $\sigma$, $\omega$, and $\rho$ mesons up to the quartic order so that the parameters should obey the naturalness behavior as imposed by the effective field theory \cite{Furnstahl1997}. 
In this work, the new parameter set is searched in view of PREX-II data and the model  EOS satisfies the observed astrophysical constraints imposed by  NSs. \\
The paper is organized as follows, in section \ref{tm}, the theoretical
framework which is used to construct the EOS for neutron stars has been discussed.
In section \ref{oaca}, the procedure for optimization and covariance analysis of the parameters is discussed. In section \ref{results}, we present our results. Finally, we summarized the results of the present work in section \ref{summary}.

\section{THEORETICAL MODEL}\label{tm}
The effective lagrangian density for the RMF model generally describes the interaction of the baryons via the exchange of $\sigma$, $\omega$, and $\rho$ mesons up to the quartic order. The Lagrangian  density\cite{Dhiman2007,Virender2022a,Raj2006} is given by
\begin{eqnarray}
\label{eq:lbm}
{\cal L} &=& \sum_{B} \overline{\Psi}_{B}[i\gamma^{\mu}\partial_{\mu}-
(M_{B}-g_{\sigma B} \sigma)-(g_{\omega B}\gamma^{\mu} \omega_{\mu}\nonumber\\&+&
\frac{1}{2}g_{\mathbf{\rho}B}\gamma^{\mu}\tau_{B}.\mathbf{\rho}_{\mu})]\Psi_{B}
+ \frac{1}{2}(\partial_{\mu}\sigma\partial^{\mu}\sigma-m_{\sigma}^2\sigma^2)\nonumber\\  &-&
\frac{\overline{\kappa}}{3!}
g_{\sigma N}^3\sigma^3-\frac{\overline{\lambda}}{4!}g_{\sigma N}^4\sigma^4  - \frac{1}{4}\omega_{\mu\nu}\omega^{\mu\nu}
+ \frac{1}{2}m_{\omega}^2\omega_{\mu}\omega^{\mu}\nonumber\\&+& \frac{1}{4!}\zeta g_{\omega N}^{4}(\omega_{\mu}\omega^{\mu})^{2}-\frac{1}{4}\mathbf{\rho}_{\mu\nu}\mathbf{\rho}^{\mu\nu}+\frac{1}{2}m_{\rho}^2\mathbf{\rho}_{\mu}\mathbf{\rho}^{\mu}\nonumber\\
&+&\frac{1}{4!}\xi g_{\rho N}^{4}(\mathbf{\rho}_{\mu}\mathbf{\rho}^{\mu})^{2} \nonumber\\
&+&  g_{\sigma N}g_{\omega N}^2\sigma\omega_{\mu}\omega^{\mu} \left(a_{1}+\frac{1}{2}a_{2}\sigma\right)\nonumber\\
&+&g_{\sigma N}g_{\rho
N}^{2}\sigma\rho_{\mu}\rho^{\mu} \left(b_{1}+\frac{1}{2}b_{2}
\sigma\right)\nonumber\\&+&\frac{1}{2}c_{1}g_{\omega N}^{2}g_{\rho N}^2\omega_{\mu}\omega^{\mu}\rho_{\mu}\rho^{\mu}
\end{eqnarray}
The equation of motion for baryons, mesons, and photons can be derived from the Lagrangian
density defined in Eq.(\ref{eq:lbm}). The equation of motion for baryons can be given as,

\begin{eqnarray}
\label{eq:dirac}
&\bigg[\gamma^\mu\left(i\partial_\mu - g_{\omega B}\omega_\mu-\frac{1}{2}g_{\rho
B}\tau_{B}.\rho_\mu - e \frac{1+\tau_{3B}}{2}A_\mu \right) - \nonumber\\
&  (M_B + g_{\sigma B}\sigma )\bigg]\Psi_B
=\epsilon_B \Psi_B.
\end{eqnarray}
The Euler-Lagrange equations for the ground-state expectation values of the mesons fields are 

\begin{eqnarray}
\label{eq:sigma}
\left(-\Delta + m_{\sigma}^{2}\right)\sigma & = &\sum_{B} g_{\sigma B}\rho_{sB} -\frac{ \overline{\kappa}}{2} g_{\sigma N}^{3}\sigma^{2}- \frac{\overline{\lambda}}{6} g_{\sigma N}^{4}\sigma ^{3} \nonumber \\ &&+ 
{a_{1}} g_{\sigma N} g_{\omega N}^{2}\omega ^{2}
 +a_{2} g_{\sigma N}^{2}
 g_{\omega N}^{2}\sigma\omega ^{2}
 \nonumber\\&&+ b_{1} g_{\sigma N} g_{\rho B}^{2}\rho ^{2}
 + b_{2} g_{\sigma N}^{2} 
g_{\rho N}^{2}\sigma\rho ^{2}, 
\end{eqnarray}
\begin{eqnarray}
\label{eq:omega}
\left(-\Delta + m_{\omega}^{2}\right)\omega & = &\sum_{B} g_{\omega B}\rho_{B} 
- \frac{\zeta}{6} g_{\omega N}^{4}\omega ^{3} 
\nonumber\\&&-2 a_{1}g_{\sigma N} g_{\omega N}^{2}\sigma\omega 
-a_{2} g_{\sigma N}^{2}
 g_{\omega N}^{2}\sigma^{2}\omega \nonumber\\
&& - c_{1} g_{\omega N}^{2} 
g_{\rho N}^{2}\omega\rho ^{2}, 
\end{eqnarray}
\begin{eqnarray}
\label{eq:rho}
\left(-\Delta + m_{\rho}^{2}\right)\rho & = &\sum_{B} g_{\rho B}\tau_{3B}\rho_{B}- \frac{\xi}{6} g_{\rho N}^{4}\rho ^{3} 
\nonumber\\ &&
-2 b_{1}
 g_{\sigma N} g_{\rho N}^{2}\sigma\rho 
- b_{2} g_{\sigma N}^{2}
 g_{\rho N}^{2}\sigma^{2}\rho \nonumber\\
&& - c_{1} g_{\omega N}^{2} 
g_{\rho N}^{2}\omega^{2}\rho,  
\end{eqnarray}
\begin{equation}
\label{eq:photon}
-\Delta A_{0} = e\rho_{p}.
\end{equation}
where the baryon vector density $\rho_B$,  scalar density $\rho_{sB}$ and charge density
$\rho_{p}$ are, respectively, 
\begin{equation}
\rho_{B}= \left< \overline{\Psi}_B \gamma^0 \Psi_B\right> = \frac{\gamma k_{B}^{3}}{6\pi^{2}},
\end{equation}

\begin{equation}
\rho_{sB} = \left< \overline{\Psi}_B\Psi_B \right> 
          = \frac{\gamma}{(2\pi)^3}\int_{0}^{k_{B}}d^{3}k \frac{M_{B}^*}
            {\sqrt{k^2 + M_{B}^{*2}}},
\end{equation}
\begin{equation}
\rho_{p} = \left< \overline{\Psi}_B\gamma^{0}\frac{1+\tau_{3B}}{2}\Psi_B \right>, 
\end{equation}
 with $\gamma$  the spin-isospin degeneracy. The $M_{B}^{*} = M_{B} - g_{\sigma B}\sigma $
is the effective mass of the baryon species B,  $k_{B}$ is its Fermi momentum
and $\tau_{3B}$ denotes the isospin projections of baryon B.
The energy density of the uniform matter  within the framework of the RMF model is given by;
\begin{equation}
\label{eq:eden}
\begin{split}
{\cal E} & = \sum_{j=B,\ell}\frac{1}{\pi^{2}}\int_{0}^{k_j}k^2\sqrt{k^2+M_{j}^{*2}} dk\\
&+\sum_{B}g_{\omega B}\omega\rho_{B}
+\sum_{B}g_{\rho B}\tau_{3B}\rho_{B}\rho
+ \frac{1}{2}m_{\sigma}^2\sigma^2\\
&+\frac{\overline{\kappa}}{6}g_{\sigma N}^3\sigma^3
+\frac{\overline{\lambda}}{24}g_{\sigma N}^4\sigma^4
-\frac{\zeta}{24}g_{\omega N}^4\omega^4\\
&-\frac{\xi}{24}g_{\rho N}^4\rho^4
 - \frac{1}{2} m_{\omega}^2 \omega ^2
-\frac{1}{2} m_{\rho}^2 \rho ^2\\
&-a_{1} g_{\sigma N}
 g_{\omega N}^{2}\sigma \omega^2
 -\frac{1}{2}a_{2} g_{\sigma N}^2 g_{\omega N}^2\sigma^2 \omega^2\\
 &-b_{1}g_{\sigma N}g_{\rho N}^2 \sigma\rho^2
-\frac{1}{2} b_{2} g_{\sigma N}^2 g_{\rho N}^2\sigma^2\rho^2\\
 &- \frac{1}{2} c_{1} g_{\omega N}^2 g_{\rho N}^2
\omega^2\rho^2.\\
\end{split}
\end{equation}
The pressure of the uniform matter  is given by
\begin{equation}
\label{eq:pden}
\begin{split}
P & = \sum_{j=B,\ell}\frac{1}{3\pi^{2}}\int_{0}^{k_j}
\frac{k^{4}dk}{\sqrt{k^2+M_{j}^{*2}}} 
- \frac{1}{2}m_{\sigma}^2\sigma^2\\
&-\frac{\overline{\kappa}}{6}g_{\sigma N}^3\sigma^3 
-\frac{\overline{\lambda}}{24}g_{\sigma N}^4\sigma^4
+\frac{\zeta}{24}g_{\omega N}^4\omega^4\\
&+\frac{\xi}{24}g_{\rho N}^4\rho^4
  + \frac{1}{2} m_{\omega}^2 \omega ^2
+\frac{1}{2} m_{\rho}^2 \rho ^2 \\
& +a_{1} g_{\sigma N}
g_{\omega N}^{2}\sigma \omega^2
+\frac{1}{2} a_{2} g_{\sigma N}^2 g_{\omega N}^2\sigma^2 \omega^2\\
&+b_{1}g_{\sigma N}g_{\rho N}^2 \sigma\rho^2
+\frac{1}{2} b_{2} g_{\sigma N}^2 g_{\rho N}^2\sigma^2\rho^2\\
&+ \frac{1}{2} c_{1} g_{\omega N}^2 g_{\rho N}^2
\omega^2\rho^2.\\
\end{split}
\end{equation}
Here, the sum is taken over nucleons and leptons.
\section{OPTIMIZATION AND COVARIANCE ANALYSIS} \label{oaca}
The optimization of the parameters ($\textbf{p}$) appearing in the Lagrangian (Eq. \ref{eq:lbm}) has been performed  by using the simulated annealing method (SAM) \cite{Burvenich2004, 
Kirkpatrick1984} by following $\chi^{2}$ minimization procedure  which is given
as, 
\begin{equation} 
	{\chi^2}(\textbf{p}) =  \frac{1}{N_d - N_p}\sum_{i=1}^{N_d}
\left (\frac{ M_i^{exp} - M_i^{th}}{\sigma_i}\right )^2 \label {chi2},
\end{equation}
where $N_d$ is the number of  experimental data points and $N_p$ is the number 
of fitted parameters. The $\sigma_i$ denotes adopted errors \cite{Dobaczewski2014}
and $M_i^{exp}$ and $M_i^{th}$ are the experimental and the corresponding
theoretical values, respectively, for a given observable.  The minimum value of 
${{\chi}}^{2}_{0}$  corresponds to the optimal values $\bf{p}_{0}$ of the parameters. 
Following the optimization of the energy density functional, it is important to explore the richness of the covariance analysis. It enables one to calculate the statistical uncertainties on model parameters or any calculated physical observables. The covariance analysis  also provides additional information about the sensitivity of the parameters to the physical observables, and interdependence among the parameters \cite{Dobaczewski2014,Chen2015,Mondal2015,Fattoyev2011}.
Having obtained the  parameter set , the correlation coefficient 
between two quantities Y and Z can be calculated by covariance analysis 
\cite{Brandt1997,Reinhard2010,Fattoyev2011,Dobaczewski2014,Mondal2015} as 
\begin{equation}
\label{covariance}
    \textit{c}_{YZ} = \frac{\overline{\Delta{Y} \Delta{Z} }}{\sqrt{\overline{\Delta{Y^2}} 
  \quad \overline{\Delta{Z^2}}}} ,
\end{equation}
where covariance between Y and Z is expressed as
\begin{equation}
\label{error}
    \overline{\Delta{Y}\Delta{Z}} = \sum_{\alpha\beta} \left( \frac{\partial{Y}}
{\partial{p}_{\alpha}}\right) _{\textbf{p}_0} C_{\alpha\beta}^{-1}  
\left( \frac{\partial{Z}}{\partial{p}_{\beta}}\right) _{\textbf{p}_0}.
\end{equation}
Here, $C_{\alpha\beta}^{-1}$ is an element of inverted curvature matrix given by
\begin{equation}
\label{matrix}
    \textit{C}_{\alpha\beta} = \frac{1}{2}\left(\frac{\partial^2 
	\chi^2(\textbf{p})}{\partial{p}_{\alpha}\partial{p}_{\beta}}\right)_{\textbf{p}_{0}}.
\end{equation}
The standard deviation, $\overline{\Delta{Y}^2}$, in Y can be computed using 
Eq. (\ref{error}) by substituting Z = Y.
 The prediction of maximum mass around $2M_{\odot}$ for the nonrotating neutron star and constraints on EOSs of Symmetric Nuclear Matter (SNM) and Pure Neutron Matter (PNM) as extracted from the analysis of particle flow in heavy ion collisions \cite{Danielewicz2002} require relatively softer EOSs as demanded by  GW170817 event. 
\begin{table*}
\centering
\caption{\label{tab:table1}
New parameter set for  DBHP model of RMF  Lagrangian given
in Eq.(\ref{eq:lbm}) along with theoretical uncertainties.
The parameters $\overline{\kappa}$, $a_{1}$, and $b_{1}$ are
 in  fm$^{-1}$. The masses  $m_{\sigma}$,$m_{\omega}$ and $m_{\rho}$  values are in  MeV. The mass for nucleon is taken as $M_N=939 MeV$.
The values of $\overline{\kappa}$, $\overline{\lambda}$, $a_{1}$, $a_{2}$, $b_{1}$, 
${b_{2}}$, 
and ${c_{1}}$ are multiplied by $10^{2}$. Parameters for NL3, FSUGarnet, IOPB-1, and Big Apple are also shown for comparison.} 
\vskip 1cm
\begin{tabular}{ccccccc}
\hline
\hline
\multicolumn{1}{c}{${\bf {Parameters}}$}&
\multicolumn{1}{c}{{\bf DBHP}}&
\multicolumn{1}{c}{{\bf NL3}}&
\multicolumn{1}{c}{{\bf FSUGarnet}}&
\multicolumn{1}{c}{{\bf IOPB-1}}&
\multicolumn{1}{c}{{\bf Big Apple}}\\
\hline
${\bf g_{\sigma}}$  &10.34155$\pm$0.06660&10.21743 &10.50315&10.41851&9.67810\\
${\bf g_{\omega}}$  &13.30826$\pm$0.10044 & 12.86762&13.69695&13.38412&12.33541\\
${\bf g_{\rho}}$ &11.25845$\pm$1.31969 &8.94800 &13.87880 &11.11560&14.14256\\
${\bf \overline {\kappa}}$&1.82166$\pm$0.05101 &1.95734 &1.65229 &1.85581&2.61776\\
${\bf \overline {\lambda}}$&0.24446$\pm$0.17154&-1.59137&-0.03533& -0.07552&-2.16586\\
${\bf{\zeta}}$&0.02156$\pm$0.00401&0.00000&0.23486& 0.01744&0.00070\\
${\bf a_1}$&0.01172$\pm$0.00383&0.00000 &0.00000 &0.00000&0.00000\\
${\bf  a_2}$  &0.05281 $\pm$0.03677&0.00000 &0.00000 &0.00000&0.00000\\
${\bf b_1}$ &0.39811$\pm$0.40926 &0.00000 &0.00000 &0.00000&0.00000\\
${\bf b_2}$ &0.09412$\pm$1.85465&0.00000 &0.00000 &0.00000&0.00000\\
${\bf c_1}$ &0.79914$\pm$3.15145&0.00000 &8.60000 &4.80000&9.40000\\
${\bf m_{\sigma}}$ &501.04834$\pm$1.34831&508.19400 &496.73100 &500.48700&492.97500\\
${\bf m_{\omega}}$&782.50000&782.50100 &782.18700 &782.18700&782.18700\\
${\bf m_{\rho}}$ &770.00000 &763.00000 &762.46800 &762.46800&762.46800\\
\hline
\hline
\end{tabular}
\end{table*}
\section{RESULTS AND DISCUSSION}\label{results}
The parameters of the model are searched by fit to the    available experimental data of total binding energies  and  charge rms radii \cite{Audinew,Otten1989,Vries1987} for some closed/open shell nuclei $^{16,24}$O,$^{40,48,54}$Ca, $^{56,68,78}$Ni,$^{88}$Sr,$^{90}$Zr,$^{100,116,132,138}$Sn,  and $^{144}$Sm, $^{208}$Pb. We have also included  the maximum mass of neutron star \cite{Rezzolla2018} in our fit data. 
\begin{figure*}
\centering
\includegraphics[trim=0 0 0 0,clip,scale=0.5]{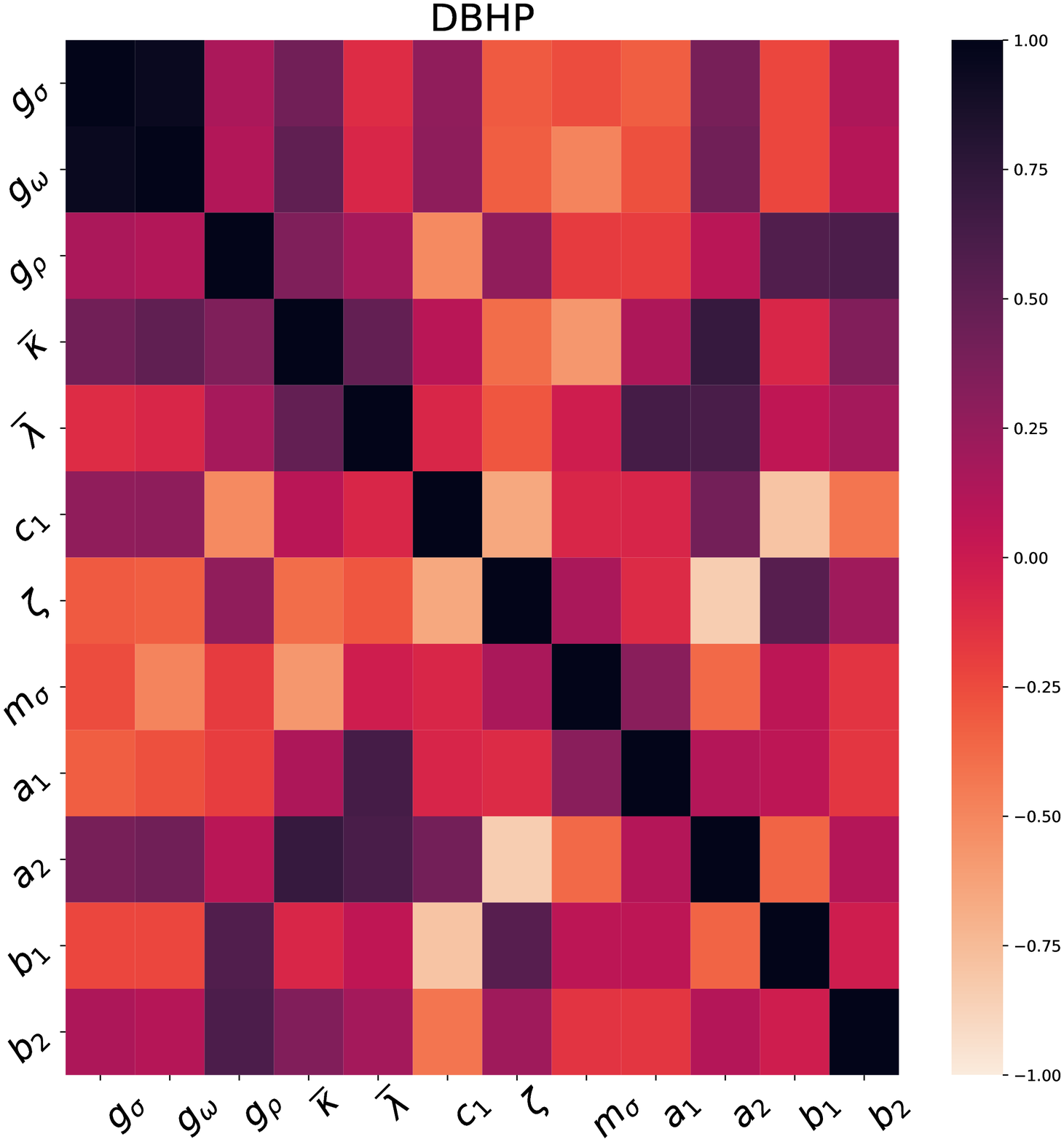}
\caption{\label{parameter_corr} (Color online) Correlation coefficients   among the model parameters for DBHP parametrization 
of the Lagrangian given by Eq. (\ref{eq:lbm}).}
\end{figure*}
Recently, the parity-violating electron scattering experiment (PREX-II) put a limit on the neutron skin thickness of $^{208}$Pb is $\Delta r_{np}$ = 0.283 $\pm$ 0.071 fm \cite{Adhikari2021}. We included the recently measured $\Delta r_{np}$ in our fit data to constrain the linear density dependence of symmetry energy coefficient.  For the open shell nuclei, the pairing has been included by using BCS formalism with constant pairing gaps that have been taken from the particle separation energies of neighboring nuclei \cite{Ring1980,Karatzikos2010,Wang2021}.
In Table \ref{tab:table1}, we display the values of relativistic parameterization DBHP  generated for the Lagrangian given by Eq. (\ref{eq:lbm}) along with theoretical uncertainties.  The values of parameter sets  for  NL3 \cite{Lalazissis1997}, FSUGarnet \cite{Chen2015}, IOPB-1 \cite{Bharat2018}  and Big Apple \cite{Fattoyev2020}  are also shown.  \\
The effective field theory imposes the condition of naturalness   \citep{Furnstahl1997} on the parameters or expansion coefficients appearing in the effective Lagrangian density  Eq. (\ref{eq:lbm}). According to naturalness, the coefficients of various terms in Lagrangian density functional should be of the same size when expressed in an appropriate dimensionless ratio. The dimensionless ratios are obtained by dividing Eq. (\ref{eq:lbm}) by $M^{4}$ and expressing each term in powers of $\frac{g_{\sigma}\sigma}{M}$, $\frac{g_{\omega}\omega}{M}$ and 2$\frac{g_{\rho}\rho}{M}$. This means that the dimensionless ratios ${\frac{1}{2C_{\sigma}^{2}M^{2}}}$,${\frac{1}{2C_{\omega}^{2}M^{2}}}$, ${\frac{1}{8C_{\rho}^{2}M^{2}}}$, ${\frac{\overline{\kappa}}{6M}}$, ${\frac{\overline{\lambda}}{24}}$, ${\frac{{\zeta}}{24}}$, ${\frac{a_1}{M}}$, ${\frac{a_2}{2}}$, ${\frac{b_1}{4M}}$, ${\frac{b_2}{8}}$ and ${\frac{c_1}{8}}$ should be roughly of same size, where ${c_{i}}^{2}=\frac{{g_{i}}^{2}}{{M_{i}}^{2}}$, i denotes $\sigma$, $\omega$ and $\rho$ mesons.
In Table \ref{tab:nat}, we present the overall naturalness behavior of DBHP parameterization i.e. the value of these parameters when expressed in dimensionless ratios as shown just above. We also display the corresponding values for NL3, FSUGarnet, IOPB-1, and Big Apple parameter sets. It is obvious from the table that DBHP  parameterization closely favors the naturalness behavior. This may be attributed to the fact that this parameterization includes all possible self and crossed interaction terms of $\sigma$, $\omega$, and $\rho$-mesons up to the quartic order. \\
 The small value of parameter $c_{1}$ for DBHP model which gives rise to better naturalness behaviour of the parameters might be attributed to the fact that the coupling parameter $c_{1}$ has strong correlation with $b_{1}$ and also has good  correlation with $a_{2}$ and $b_{2}$ (see Fig. \ref{parameter_corr}).
It is evident from the table that the value of coupling parameter $c_{1}$ (crossed interaction term of $\omega^{2}$ and $\rho^{2}$) appearing in Eq. (\ref{eq:lbm})  is large  for IOPB-I, FSU-Garnet and Big Apple which  shows deviation from the naturalness behavior  in the absence of  all other  possible mixed interaction terms of $\sigma$, $\omega$, and $\rho$-meson.
\begin{table*}
\centering
\caption{\label{tab:nat}
The values of parameters are expressed as dimensionless ratios corresponding to naturalness behavior. All  values have been  multiplied by $10^{3}$.} 
\begin{tabular}{cccccccc}
\hline
\hline
\multicolumn{1}{c}{{ {Parameters}}}&
\multicolumn{1}{c}{{ DBHP}}&
\multicolumn{1}{c}{{ NL3}}&
\multicolumn{1}{c}{{ FSUGarnet}}&
\multicolumn{1}{c}{{IOPB-1}}&\multicolumn{1}{c}{{ Big Apple}}\\
\hline
${\bf{\frac{1}{2C_{\sigma}^{2}M^{2}}}}$ & 1.3311&1.4028 &1.2690&1.3086&1.4698\\
${\bf{\frac{1}{2C_{\omega}^{2}M^{2}}}}$  &1.9604& 2.0970&1.8508&1.9383&2.2819\\
${\bf{\frac{1}{8C_{\rho}^{2}M^{2}}}}$ &0.6631 &1.0306&0.4278&0.6670 &0.4121\\
${\bf{\frac{\overline{\kappa}}{6M}}}$&0.6380 &0.6855 &0.5787 &0.6499&0.9168\\
${\bf{\frac{\overline{\lambda}}{24}}}$&0.1018&-0.6630&-0.1472& -0.3146&-0.9024\\
${\bf{\frac{{\zeta}}{24}}}$&0.8982&-&0.9785 & 0.7267&0.0291\\
${\bf{\frac{a_1}{M}}}$ &0.1172 &- &- &-&-\\
${\bf{\frac{a_2}{2}}}$&0.2641 &- &- &-&-\\
${\bf{\frac{b_1}{4M}}}$ &0.9953 &- &- &-&-\\
${\bf{\frac{b_2}{8}}}$ &0.1177 &- &- &-&-\\
${\bf{\frac{c_1}{8}}}$ &0.9989 &- &10.7500 &6.0000&11.7500\\
\hline
\hline
\end{tabular}
\end{table*}
\begin{table*}
\caption{\label{tab:table2}
The bulk nuclear matter properties (NMPs) at saturation density along with calculated theoretical errors for DBHP parameterization compared with that other parameter sets. $\rho_{0}$,  E/A,  K, $M^{*}/M$, J,  L and $K_{sym}$  denote the saturation density, Binding Energy per nucleon, Nuclear Matter incompressibility coefficient, the ratio of effective nucleon mass to the nucleon mass, Symmetry Energy, the slope of symmetry energy, and curvature of symmetry energy   respectively. the value of $\rho_0$ is in fm$^{-3}$ and rest all the quantities are in MeV. The values of neutron skin thickness $\Delta r_{np}$ for $^{208}$Pb and $^{48}$Ca nuclei in units of fm are also listed.}
\vskip 1cm
\begin{tabular}{cccccc}
\hline
\hline
\multicolumn{1}{c}{${ {NMPs}}$}&
\multicolumn{1}{c}{{ DBHP}}&
\multicolumn{1}{c}{{ NL3}}&
\multicolumn{1}{c}{{ FSUGarnet}}&
\multicolumn{1}{c}{{ IOPB-1}}&
\multicolumn{1}{c}{{ Big Apple}}\\
\hline
${\bf{\rho_{0} }}$&0.148$\pm$0.003&0.148 &0.153&0.149&0.155\\
${\bf E/A}$ &-16.11$\pm$0.05 & -16.25&-16.23&-16.09&-16.34\\
${\bf K}$ &229.5$\pm$5.6&271.6 &229.6&222.6&227.1\\
${\bf ~M^{*}/M}$&0.615$\pm$0.007&0.595&0.578& 0.595&0.608\\
${\bf J}$&34.7 $\pm$1.5&37.4&30.9&33.3&31.4\\
${\bf L}$&83.9$\pm$19.2&118.6&50.9& 63.8&40.3\\
{${\bf {K_{sym}} }$}&-33.2$\pm$64.1&100.7&57.9& -38.4&88.8\\
{${\bf {\Delta r_{np}~ (^{208}Pb)} }$}&0.24$\pm$0.02&0.28&0.16&0.22&0.15\\
${\bf {\Delta r_{np} ~(^{48}Ca)}}$&0.21$\pm$0.02&0.23&0.17&0.17&0.17\\
\hline
\hline
\end{tabular}
\end{table*}
  Keeping in view the naturalness behavior of the parameters as imposed by the effective field theory \cite{Furnstahl1997} and as observed  in case of DBHP model, it is important to incorporate the  contributions of the higher order mixed interactions of mesons in the Lagrangian. The naturalness behavior of parameters can be further improved by considering the next higher order terms containing the gradient of fields   \citep{Furnstahl1997}. As far as NL3 parameterization is concerned, the naturalness behavior is favored very well but it does not include any cross interaction terms of $\sigma$, $\omega$, and $\rho$ mesons which are very important for constraining the symmetry energy and its density dependence. \\
In Fig.\ref{parameter_corr}, the correlation coefficients between the DBHP model parameter appearing in Lagrangian (Eq. \ref{eq:lbm}) are shown in graphical form. A strong correlation is found between the pairs of model parameters $g_{\sigma}$ and $g_{\omega}$ (0.95), $c_{1}- b_{1}$ (0.80), and $a_{2}-\overline{\kappa}$ (0.72). The strong  correlation is also found for $g_{\rho}$  with $b_{1}$ and $b_{2}$.  Mild correlations  exist   between the pairs of model parameters $g_{\sigma} -$ $\overline{\kappa}$, $g_{\sigma}-$ $a_{1}$ and $g_{\sigma}-$ $a_{2}$. A strong correlation between the model parameters implies a strong interdependence i.e. if one parameter is fixed at a certain value then the other must attain the precise value as suggested by their correlation. \\
\subsection{Properties of finite nuclei and nuclear matter }
\begin{figure}
\includegraphics[trim=0 0 0 0,clip,scale=0.45]{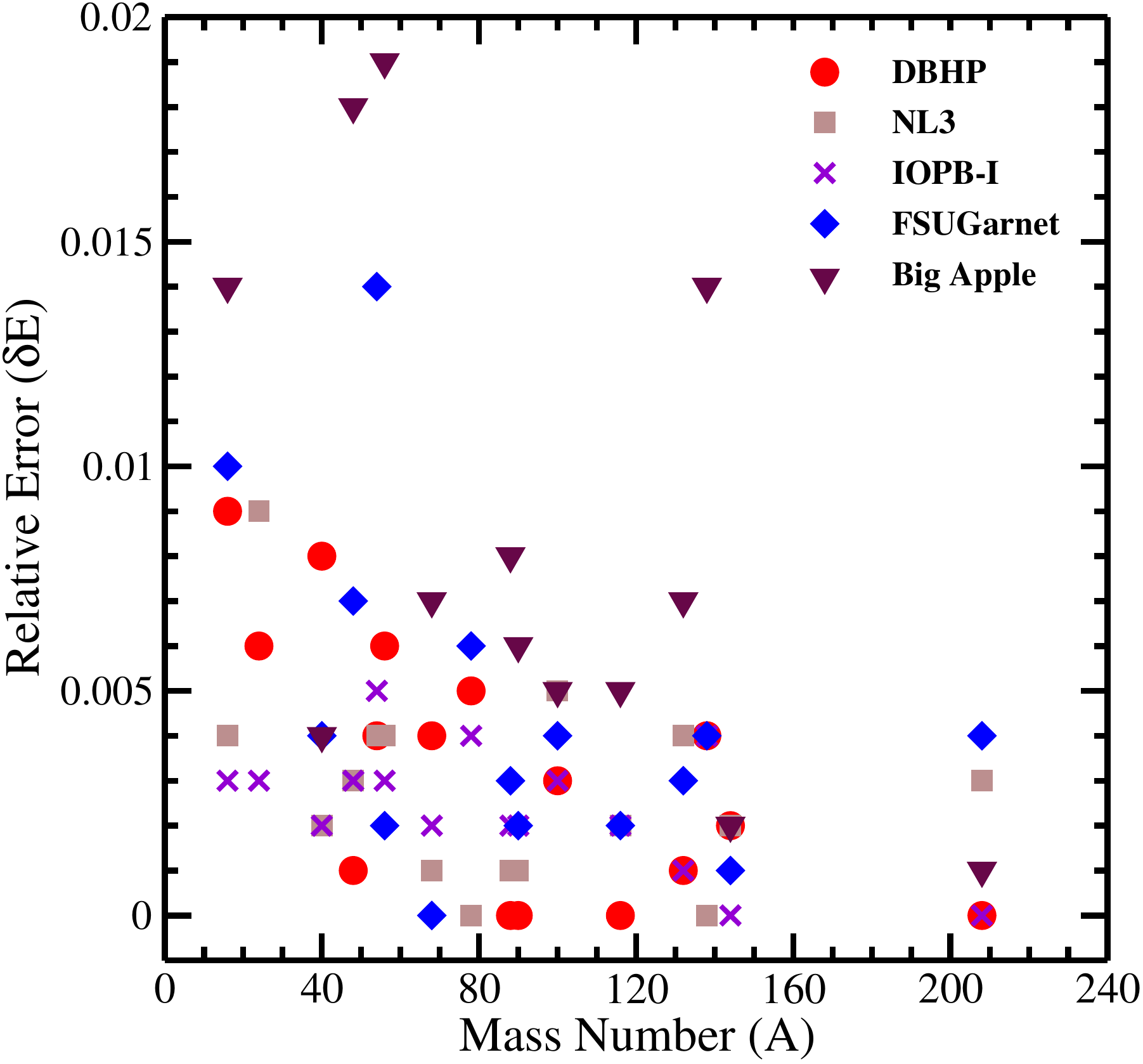}
\caption{\label{error_be} (Color online) Relative error in the total binding energy ($\delta E$) plotted against the mass number (A) for the newly generated parameter set DBHP. For comparison, the values of $\delta E$ obtained with parameters NL3, IOPB-1, FSUGarnet and Big Apple are also displayed. }
\end{figure}

\begin{figure}
\includegraphics[trim=0 0 0 0,clip,scale=0.58]{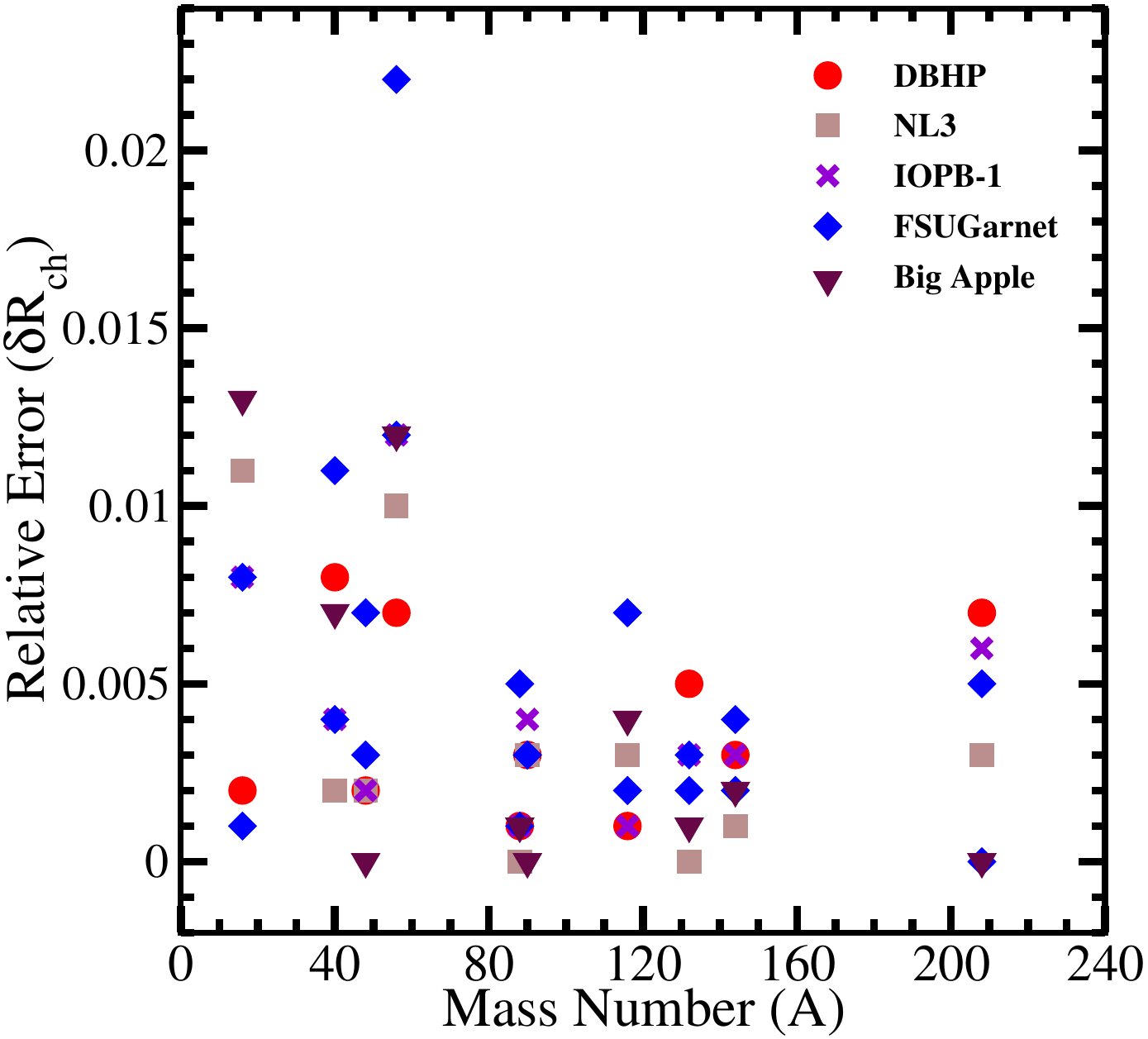}
\caption{\label{error_ch} (Color online) Relative error in the charge root mean square ($\delta R_{ch}$) plotted against the mass number (A) for the newly generated parameter set DBHP For comparison, the values obtained with parameters NL3, IOPB-1, FSUGarnet and Big Apple are also displayed.}
\end{figure}

The newly generated DBHP parameterization gives a good fit to the properties of finite nuclei.  In Fig. \ref{error_be}, we display the value of relative error in the total binding energies $\delta E = \frac{B^{exp}-B^{th}}{B^{exp}}$ calculated for DBHP parameterization. We also display similar results for  other  parameter sets considered. It is evident that binding energies obtained using DBHP parameterization are in good agreement with the available experimental data   \citep{Wang2017}. The root mean square (rms) errors in total binding energy for all the nuclei considered in our fit data is found to be 2.1 MeV. In Fig.\ref{error_ch}, we present our results for relative error $\delta R_{ch}$ for charge rms radii and also compare them with  other  parameter sets. The root mean square (rms) errors in  charge radii for all nuclei taken in our fit is 0.02 fm. The neutron skin thickness of $^{208}$Pb for DBHP model comes out to be 0.24 $\pm$ 0.02 fm.
In Table \ref{tab:table2}, we present the results for the  SNM properties such as binding energy per nucleon (E/A), incompressibility (K), the  effective nucleon mass ($M^*$)   at the saturation density ($\rho_{0}$),   symmetry energy coefficient (J), slope of  symmetry energy (L) and curvature parameter  $K_{\rm sym}$   along with the theoretical uncertainties.  It is observed that the isoscalar properties (E/A, K, $M^*$ $\rho_{0}$) are well constrained for DBHP parametrization (at the $\leq $ 3.3 \% level). But in isovector sector, the  error on the density dependence of the symmetry energy are relatively larger  for $L$ ($\approx$ 23 \%). The value of $K_{\rm sym}$ is determined only poorly \cite{Newton2021,Xu2022,Gil2022}.  
The experimental data on finite nuclei are not enough to  constrain $K_{\rm 
sym}$. Only the accurate knowledge of symmetry energy at higher densities
($\rho > 2\rho_0$)   may constrain the $K_{\rm sym}$ in tighter bounds.
This may be attributed to the large experimental error on the neutron skin thickness for  $^{208}$Pb (0.283 $\pm$ 0.071 fm) which lead us choosing the large adopted error during the optimisation procedure. The values of neutron-skin thickness $(\Delta r_{np})$ for $^{208}$Pb and $^{48}$Ca nuclei are also presented.  The DBHP parameter significantly overestimates the value of  neutron-skin
thickness  for $^{48}$Ca  in comparison to that  $\Delta r_{np} (^{48}{\rm
Ca})=0.121\pm 0.026 $ fm as reported recently  by CREX \cite{Adhikari2022}.  Other parametrizations considered
in Table \ref{tab:table2} also do not satisfy simulteneously  the
experimental data for the neutron-skin for the $^{208}$Pb and $^{48}$Ca
nuclei. Similar trends have been observed in recent investigations based
on the  relativistic and non-relativistic mean field models which call
for  further experimental studies
\cite{Paar2022,Reinhard2022,Piekarewicz2022}.
\begin{figure}
\includegraphics[trim=0 0 0 0,clip,scale=0.48]{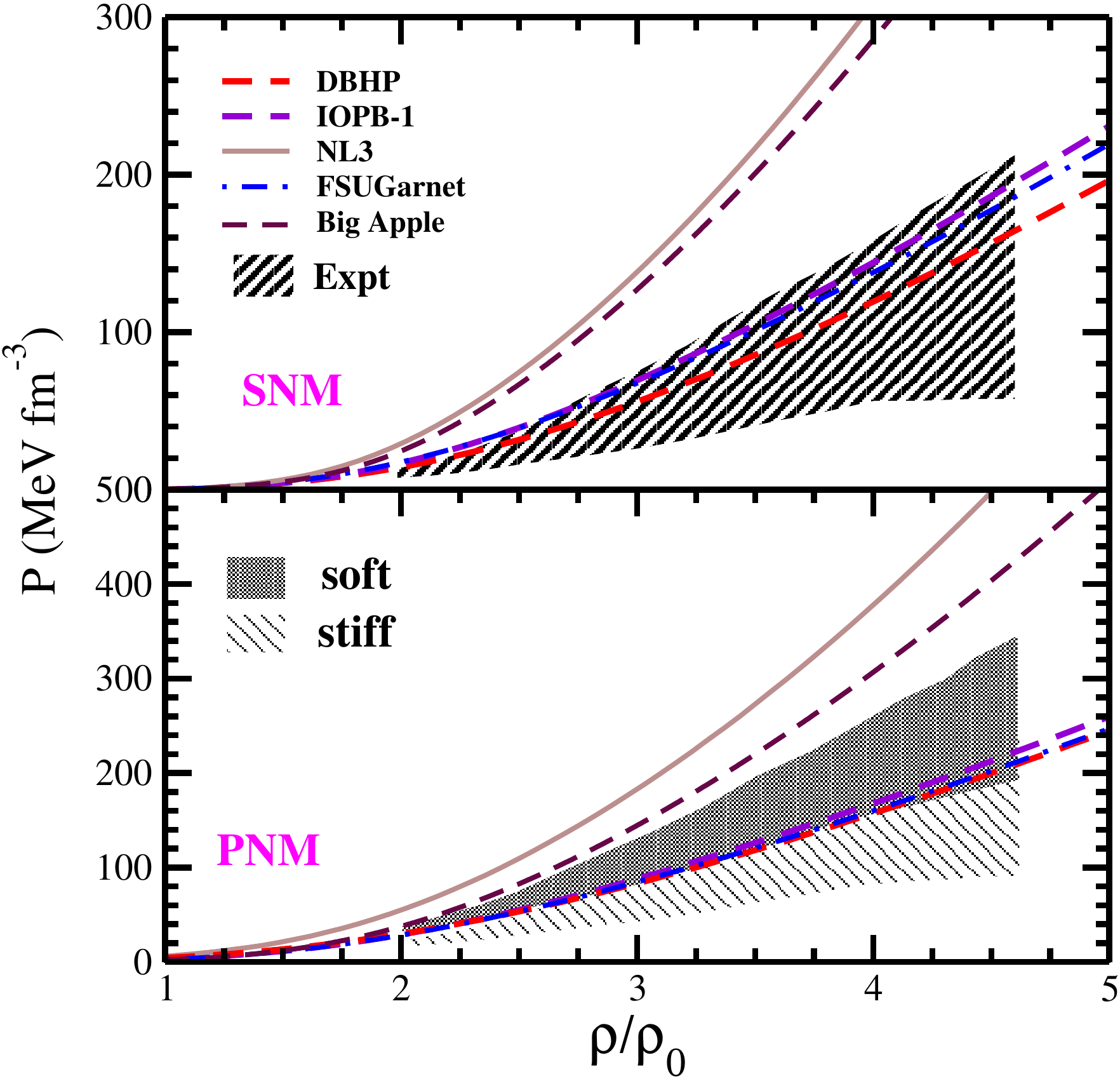}
\caption{\label{snm_pnm} (color online) Variation of Pressure as a function of baryon density for SNM (upper panel) and PNM (lower panel) computed with  DBHP parameterization along with NL3, IOPB-1, FSUGarnet and Big Apple models. The shaded region represents the experimental data taken from the reference \cite{Danielewicz2002}.}
\end{figure}

\begin{figure}
\centering
\includegraphics[trim=0 0 0 0,clip,scale=0.5]{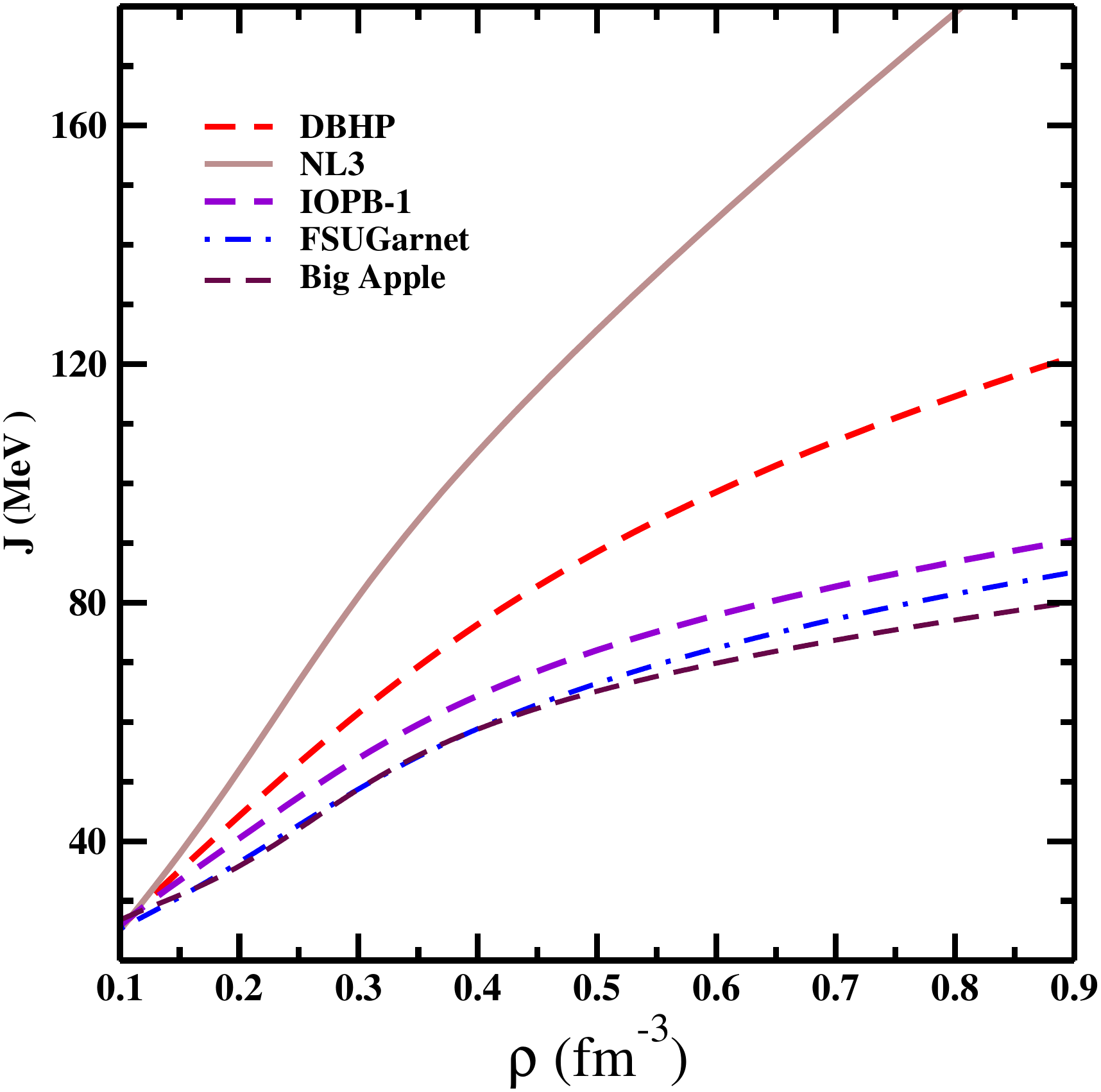}
\caption{ \label{sym} (Color online) The density dependent  symmetry energy  plotted as a function of baryon density for DBHP model. The results are also displayed for NL3, IOPB-1, FSUGarnet and Big Apple parameter sets.}
\end{figure}

The results  are also compared with the NL3 \cite{Lalazissis1997}, FSUGarnet \cite{Chen2015}, IOPB-1 \cite{Bharat2018}  and Big Apple \cite{Fattoyev2020} parameter sets. These SNM properties are very important for constructing the EOS for nuclear matter. E/A is -16.1 MeV for DBHP parameterization. The  value of J and L obtained by DBHP parameterization are  consistent with the values J = 38.1 $\pm$ 4.7 MeV and L = 106 $\pm$ 37 MeV as  inferred by Reed et. al., \cite{Reed2021}. The value of K is 225 MeV  which is in agreement with the value K = 240 $\pm$ 20 MeV determined from isoscalar giant monopole resonance (ISGMR) for  $^{90}$Zr and $^{208}$Pb nuclei       \citep{Colo2014,Piekarewicz2014}.  \\
In Fig. \ref{snm_pnm}, we plot the EOS i.e. pressure as a function of the baryon density for SNM (upper) and  PNM (lower panel) using the DBHP parametrization that agrees reasonably well and lies in the allowed region with the EOS extracted from the analysis of the particle flow in heavy ion collision \cite{Danielewicz2002}.
It is evident from the figure that the EOSs for SNM and PNM calculated with the NL3 parameterization are very stiff and ruled out by the heavy ion collision data. The EOS calculated by using the DBHP parameterization is relatively softer which is in requirement to constrain the recent astrophysical observations \cite{Steiner2010,Rezzolla2018,Riley2021,Miller2021}. In Fig. \ref{sym}, we plot the symmetry energy as a function of baryon density for DBHP model. The results for  other parametrizations  are also shown for comparison. It can be observed that the symmetry energy increases with  baryon density and it is found to be softer than NL3 but stiffer than IOPB-1, FSUGarnet and Big Apple models.

\subsection{Neutron star properties}
\begin{figure}
\hspace{-0.3cm}\includegraphics[trim=0 0 0 0,clip,scale=0.47]{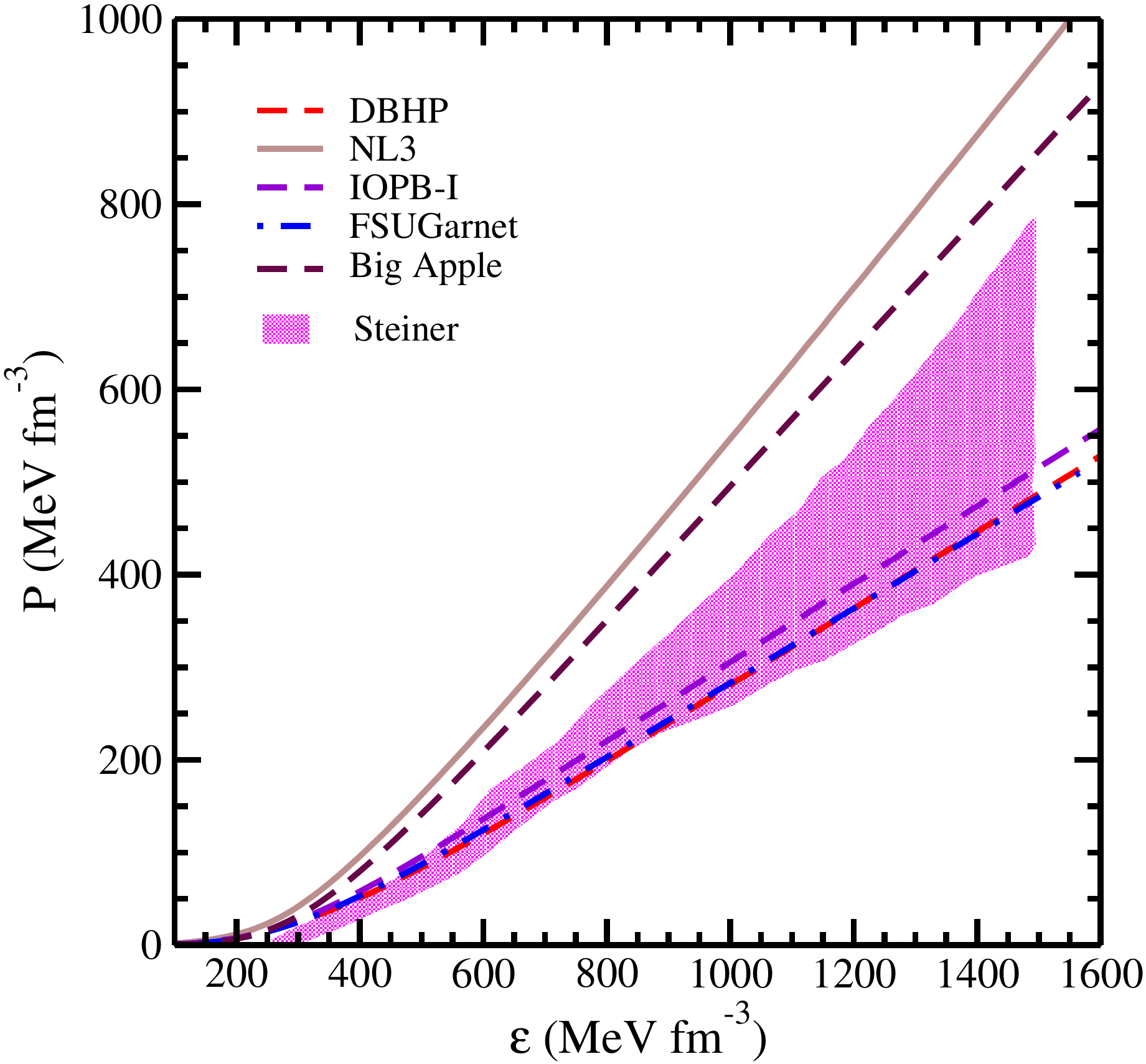}
\caption{\label{fig:fig_1} (color online) Variation of Pressure as a function of Energy Density for DBHP parameter set. EOS computed with NL3, IOPB-1, FSUGarnet and Big Apple models  are also shown for comparison. The Shades region represents the observational constraints taken from reference \cite{Steiner2010}.}
\end{figure}

 In  Fig. \ref{fig:fig_1} we display     the variation of  pressure with the energy density  for the nucleonic matter  in $\beta$ equilibrium for  the DBHP parameterization. The results are also compared with those obtained for  parameter sets. The shaded region represents the observational constraints at $r_{ph}$=R with the 2$\sigma$ uncertainty   \cite{Steiner2010}. Here $r_{ph}$ and R  are the photospheric and neutron star radius respectively. It is clear that the EOS computed with our  DBHP parameter set is consistent with the EOS obtained by Steiner et al. \cite{Steiner2010}.

\begin{figure}
\hspace{-0.3cm}\includegraphics[trim=0 0 0 0,clip,scale=0.49]{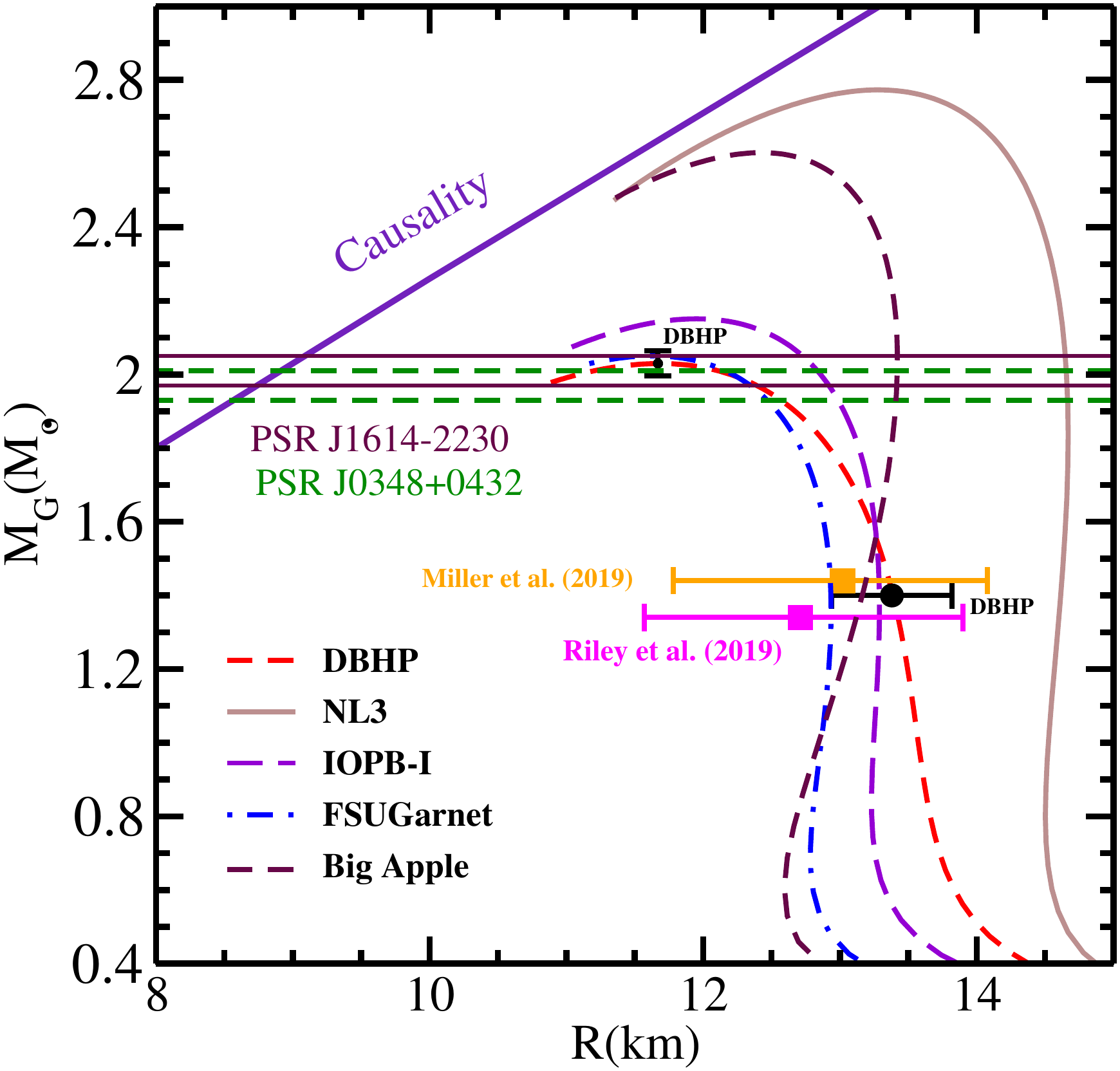}
\caption{\label{fig:fig_2} (color online) Relationship between neutron star mass and its radius for DBHP parameterization. The results are compared with NL3, IOPB-1, FSUGarnet and Big Apple parameters.}
\end{figure}

The EOSs obtained by the DBHP and IOPB-1  parameterizations are softer and lie in the allowed shaded region which represents the observational constraints taken from Ref. \cite{Steiner2010}. 
The EOS obtained with NL3 parameter set is much stiffer than DBHP and IOPB-1 parameter sets and ruled out by the observational constraints \cite{Steiner2010}. The stiffness of EOS for NL3 may be attributed to its very high value of compressibility (K), symmetry energy coefficient (J), and slope of symmetry energy (L) as shown in Table (\ref{tab:table2}). The mass and radius of a neutron star are obtained by solving the Tolman-Oppenheimer-Volkoff (TOV) equations \cite{Oppenheimer1939,Tolman1939} given as:

\begin{equation}
\label{eq:tov}
\frac{dP(r)}{dr} = -\frac{\{\epsilon(r)+P(r)\}\{4\pi r^3 P(r)+m(r)\}}{r^2(1-2m(r)/r)}
\end{equation}
\begin{equation}
\label{eq:nr31}
\frac{dm}{dr}=4\pi r^2\epsilon(r),
\end{equation}
\begin{equation}
m(r)= 4\pi\int_0^{r}dr r^2 \epsilon(r) 
\end{equation}

where $P(r)$  is the pressure at radial distance $r$ and $m(r)$  is the  mass of neutron stars  enclosed in the sphere of radius $r$. The EOS for the crust region is taken from Ref.  \cite{Sugahara1994}.
In Fig. \ref{fig:fig_2} we present our results for gravitational mass of static  neutron  star and its radius  for DBHP and other  parameterizations.

It is observed that the maximum gravitational mass of the static  neutron star for DBHP parameter set is 2.03 M$\odot$  which is in good agreement with the mass constraints from GW170817 event, pulsars PSRJ1614-2230, PSRJ0348+0432, and PSRJ0740+6620  \cite{Demorest2010,Fonseca2016,Rezzolla2018,Riley2021,Miller2021}.
\begin{table*}
 \caption{\label{tab:table4}The properties of nonrotating neutron stars along with theoretical uncertainties obtained for the DBHP parameter set. Results are also compared with the other parameter sets. $M_{\rm max}$ and $ R_{\rm max}$ denote the Maximum Gravitational mass and corresponding radius, respectively. The  values for $R_{1.4}$ and $\Lambda_{1.4}$ denote radius and  dimensionless tidal deformability at $1.4M_\odot$.}
 \begin{tabular}{ccccc}
\hline
\bf{EOS}& \bf{M}& \bf{R$_{max}$ } & \bf{R$_{1.4}$}&\bf{$\Lambda_{1.4}$}\\
 & (M$_{\odot}$) &(km) &(km) & \\
 \hline
DBHP&2.03$\pm$0.04&11.68$\pm$0.29&13.39$\pm$0.41&682$\pm$125\\
NL3&2.77&13.27&14.61&1254\\
IOPB-I&2.15&11.95&13.28&694\\
FSUGarnet&2.06&11.70&12.86&624\\
Big Apple&2.6&12.41&12.96&717\\
\hline
\end{tabular}
\end{table*}
 The radius ($R_{1.4}$) of canonical mass  is 13.39 Km for DBHP parameterization  which satisfies the radius constraints from NICER \cite{Annala2018,Riley2021,Miller2021}.  The value of  $R_{1.4}$for NL3 parameterization is 14.61 Km  which seems to rule out the constraints for $R_{1.4}$ extracted from Ref. \cite{Annala2017}.
 
\begin{figure}
\centering
\includegraphics[trim=0 0 0 0,clip,scale=0.45]{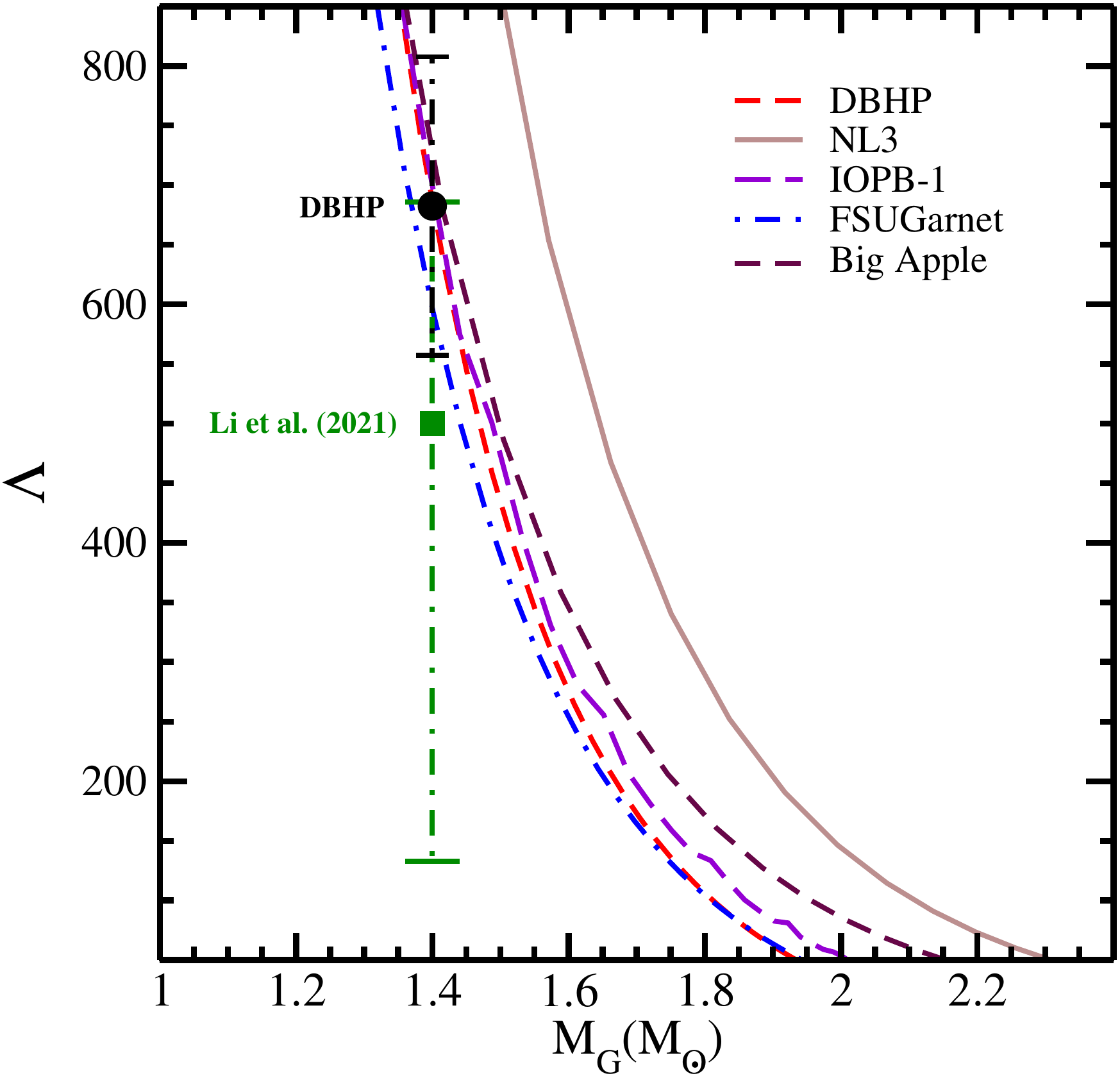}
\caption{\label{tidal} (Color online) Variation of  dimensionless tidal deformability ($\Lambda$)
 with respect
to gravitational mass for DBHP parameterization. The results for NL3, IOPB-1, FSUGarnet and Big Apple parameters are also shown.}
\end{figure}

The tidal deformability $\Lambda$ rendered by the companion stars on each other in a binary system can provide remarkable pieces of information on the EOS of neutron stars \cite{Hinderer2008,Hinderer2010}. 
The tidal influences of its companion in BNS system will deform neutron stars in the binary system and, the resulting change in the gravitational potential modifies 
the BNS orbital motion and its corresponding gravitational wave (GW) signal. This effect on GW phasing can be parameterized by the dimensionless tidal deformability parameter,
$\Lambda_i = \lambda_i/M_i^5,$ i = 1, 2.  For each neutron star, its quadrupole moment ${\cal{Q}}_{j,k}$ must be related to the tidal field ${\cal{E}}_{j,k}$ caused by its companion
as, ${\cal{Q}}_{j,k} = -\lambda {\cal {E}}_{j,k}$, where, $j$ and $k$ are spatial tensor indices.
The dimensionless tidal deformability
parameter  $\Lambda$  of a static, spherically symmetric compact star
depends on the neutron star compactness parameter C and a dimensionless quadrupole Love number k$_{2}$ as, $\Lambda = \frac{2}{3} k_2 C^{-5}$. The $\Lambda$ critically parameterizes 
the deformation of neutron stars under the given tidal field, therefore it should depend on the EOS of nuclear dense matter. 
To measure the Love number $k_{2}$ along with the evaluation of the TOV  
equations we have to compute $y_{2} = y(R)$ with 
initial boundary condition y(0) = 2 from the first-order
differential equation   \citep{Hinderer2008,Hinderer2009,Hinderer2010,Damour2010} simultaneously,
 \begin{eqnarray}
 \label{y}
   y^{\prime}&=&\frac{1}{r}[-r^2Q-ye^{\lambda}\{1+4\pi Gr^2(P-{\cal{E}})\}-y^{2}], \\
  Q &\equiv & 4\pi Ge^{\lambda}(5{\cal{E}}+9P+\frac{{\cal{E}}+P}{c_{s}^2})
 -6\frac{e^{\lambda}}{r^2}-\nu^{\prime^2}\\
   e^{\lambda} &\equiv& (1-\frac{2 G m}{r})^{-1}\\
 \nu^{\prime}&\equiv&  2G e^{\lambda} 
       (\frac{m+4 \pi P r^3}{r^2}).
      \end{eqnarray} 
First, we get the solutions of Eq.(\ref{y}) with boundary condition, y$_{2}$ = y(R),
then the electric tidal Love
number k$_{2}$ is calculated from the expression as,
\begin{eqnarray}
 k_{2}&=& \frac{8}{5}C^{5}(1-2C)^{2}[2C(y_{2}-1)-y_{2}+2]\{2C(4(y_{2}+1)C^4\nonumber\\
 &+&(6y_{2}-4)C^{3}
 +(26-22y_{2})C^2+3(5y_{2}-8)C-3y_{2}+6)\nonumber\\
 &-&3(1-2C)^2(2C(y{_2}-1)-y_{2}+2) \log(\frac{1}{1-2C})\}^{-1}.\nonumber\\
\end{eqnarray}
Fig. \ref{tidal} shows the results of dimensionless tidal deformability $\Lambda$ as a function of gravitational mass for neutron stars for DBHP and other  parameterizations. The value of  $\Lambda$ decreases with an increase in the gravitational mass of the neutron star and reduces to a very small value at the maximum mass. The value of $\Lambda_{1.4}$ obtained for canonical mass with DBHP parameters is 682 $\pm$ 125 which satisfies the finding from the GW170817 event \cite{Reed2021,Abbott2017,Chen2021} for the EOS of dense nuclear matter. \\
It is noteworthy that the our analysis of  tidal deformability  ($\Lambda_{1.4}$) lies  within  the constraint ($\Lambda_{1.4} \leq $ 800) for GW170817 event  \cite{Abbott2017}. But value of $\Lambda_{1.4}$ obtained for DBHP model (682) has marginal overlap  with revised   limit $\Lambda_{1.4} \leq  580$ within $1\sigma$ uncertainty  \cite{Abbott2018} . This is attributed to  the impact of inclusion of PREX-II data in our fit which produces stiff symmetry energy with   density slope L = 83.9 MeV.  We are looking forward that  the new terrestrial experiments and astrophysical observations may  impose  tighter  bounds.
\\

In Table \ref{tab:table4}, we present the results for the various properties of static  stars with DBHP parameterization. The theoretical uncertainties calculated for the properties using Eqs. (\ref{covariance} and \ref{error}) are also listed. Results obtained with other parameter sets are also shown for comparison.
We obtain a very small theoretical uncertainties  for the maximum mass $M_{max}$ (1.9 \%), maximum mass radius $R_{max}$ (2.5 \%) and  radius $R_{1.4}$ (3 \%) of  neutron star. The small uncertainties  might be attributed to the fact that the  inclusion of $M_{max}$ in the   fit data constraint the high density regime of EOS. A relatively large uncertainties ($\approx$ 18 \%) is obtained for $\Lambda_{1.4}$. This is due to the fact that $\Lambda \propto R^5$ which  indicates that
precise measurement of  tidal deformability  can constrain the NS radius in narrow bounds.
Indeed it is  believed that no terrestrial experiment can reliably constrain the EOS of neutron star \cite{Reed2021}.

\subsection{Correlations of nuclear matter, neutron star properties and model parameters}
\begin{figure*}
\centering
\includegraphics[trim=0 0 0 0,clip,scale=0.45]{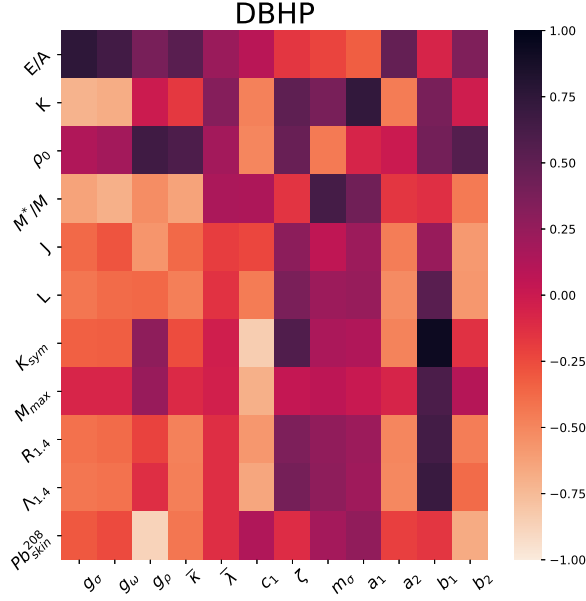}
\caption{\label{par_vs_snm_skin} (Color online) Correlation coefficients 
 between the model parameters and a set of neutron star  
	observables  as well as the  bulk properties  of nuclear matter at the saturation density for DBHP parametrization (see text for details).}
\end{figure*}

We now  discuss the correlation coefficients, shown  in Fig. \ref{par_vs_snm_skin}, between the model parameters and  nuclear matter properties, neutron skin thickness of $^{208}$Pb nucleus as well as NS  observables.
\begin{figure*}
\centering
\includegraphics[trim=0 0 0 0,clip,scale=0.6]{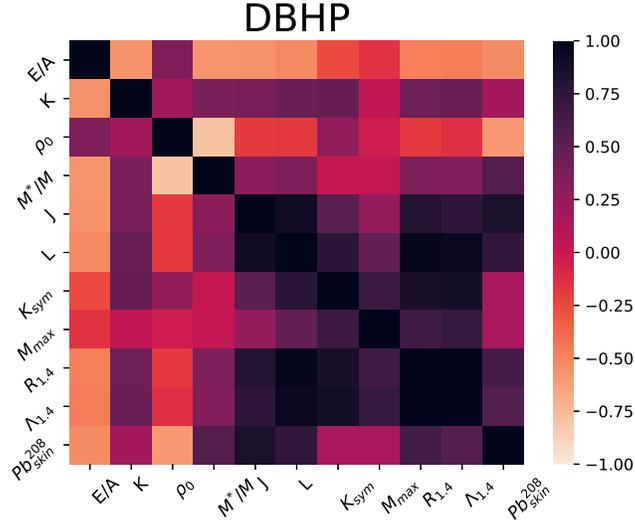}
\caption{\label{snm_skin_corr} (Color online) Correlation coefficients 
 for  bulk nuclear matter and neutron start properties and neutron 
skin of $^{208}$Pb for DBHP parametrization.}
\end{figure*}
The isoscalar nuclear matter properties like E/A, K, M*/M show strong correlations with isoscalar parameters $g_{\sigma}$, $g_{\omega}$ and $\overline{\kappa}$.  It can also be observed from the figure that the  symmetry energy  slope parameter $(L)$ can be constrained by the coupling parameter $a_{2}$, $b_{1}$ and $b_{2}$ along with the coupling parameter $g_{\rho}$ as suggested by their correlations.  The value of $\Delta r_{np}$ is found to be well constrained by the parameters   $g_{\rho}$ and $b_{2}$ as they have strong correlations.  This study is quite consistent with results reported in  \cite{Chen2015,Fattoyev2011}.
Finally, we  discuss the correlations between neutron star observables and Lagrangian model parameters as shown in Fig. \ref{par_vs_snm_skin}. A strong  correlation between maximum neutron star mass and $\omega$-meson self-coupling parameter $\zeta$ is missing in case of DBHP model parameterization. The  $M_{max}$ display a moderate  correlations  with isovector coupling parameter $c_{1}$ and $b_{1}$.
A large maximum mass may be generated either by having a stiff EOS for  SNM or a stiff symmetry energy. If the symmetry energy is soft , then one must stiffen EOS of SNM which can be done by tuning the parameter $\zeta$. But the symmetry energy of DBHP model is stiff  as shown by the  Fig. \ref{sym}.
The symmetry energy  slope  parameter at saturation density  is found to be  83.9 MeV. The stiff symmetry energy thereby weakens  the correlation between $\zeta$ and $M_{max}$. This suggests that the maximum mass results from a competition between $\zeta$ and $L$. This further implies that the parameter $\zeta$ should be well correlated to $c_{1}$ and $b_{1}$ and this is what exactly reflected from the correlations shown in Fig.\ref{parameter_corr}. The values of  $L$ and $K_{sym}$ are   found to be constrained by the parameters $c_{1}$ and $b_{1}$.
Finally, in Fig. \ref{snm_skin_corr}  we display the correlation coefficients between the properties of nuclear matter, neutron star, and neutron skin thickness of $^{208}$Pb.  A strong correlation of neutron skin thickness of $^{208}$Pb nucleus with J, L, $R_{1.4}$ and $\Lambda_{1.4}$ is observed. As per the  expectation, radius  $R_{1.4}$ is found to have a strong correlation with J and L. These findings are quite in harmony with the results reported in Ref. \cite{Fattoyev2011,Chen2015}. The curvature of the symmetry energy ($K_{sym}$) is also found to have a strong correlation with $R_{1.4}$ and $\Lambda_{1.4}$.

\section{Summary}\label{summary}
The new relativistic interaction DBHP for the relativistic mean field model has been generated by keeping in view the  PREX-II data for  neutron-skin in $^{208}$Pb nucleus, astrophysical constraints in addition to those usually employed, like, binding energy, charge radii  for fintie nuclei  and empirical data on nuclear matter at the saturation density.  We have included all possible self and mixed interactions between $\sigma$, $\omega$, and $\rho$-meson up to the quartic order so that the coupling parameters obey the naturalness behavior as imposed by the effective field theory \cite{Furnstahl1997}.  The  Covariance analysis enabled us to asses  the statistical uncertainties in the estimation of the model parameters   and   observables of interest as well as the  correlations among them.
The  DBHP parameter set is obtained such that it reproduces the ground state properties of the finite nuclei,  bulk nuclear matter properties and also satisfies the constraints of mass and radius of the neutron star and its dimensionless deformability $\Lambda$ from recent astrophysical observations \cite{Steiner2010,Annala2017,Abbott2018,Abbott2019}.
 The root mean square errors in the total binding energies and charge rms radii for finite nuclei included in our fit for DBHP parameterization are 2.1 MeV and 0.02 fm respectively. The Bulk nuclear matter properties obtained  are well consistent with the current empirical data \cite{Reed2021,Piekarewicz2014}.
 The maximum gravitational mass and radius ($R_{1.4}$) of the neutron star comes out to be  2.029 $\pm$ 0.038 M$\odot$  and  13.388 $\pm$ 0.521  km respectively \cite{Rezzolla2018}.  The value of $\Lambda_{1.4}$ which is equal to 682.497$\pm$125.090 for DBHP parameterization also satisfies the constraints for GW170817  event \cite{Abbott2017} and  reported in Refs. \cite{Chen2021,Reed2021}. The parametrization generated in consideration of PREX-II data produces stiff symmetry energy coefficient and its density depdendence leading to  the  $\Lambda_{1.4} =  682 \pm 125$  which has marginal overlap with the revised constraint \cite{Abbott2018}.  We are looking forward that  the new terrestrial experiments and astrophysical observations may put  more stringent constraints on the density dependence of the symmetry energy.

\begin{acknowledgments}
V.T. is highly thankful to Himachal Pradesh University
for providing computational facility and the Department of Science \& Technology (Govt. of India) for providing financial
assistance (DST/INSPIRE Fellowship/2017/IF170302) under
Junior/Senior Research Fellowship scheme.
 C.M. acknowledges partial support 
 from the IN2P3 Master Project “NewMAC”.
\end{acknowledgments}
  \bibliography{Ref}
  \bibliographystyle{unsrt}
\end{document}